\documentclass[reprint,
superscriptaddress,
amsmath,amssymb,
aps,
pra,
]{revtex4-2}

\usepackage{graphicx}% Include figure files
\usepackage{dcolumn}% Align table columns on decimal point
\usepackage{bm}% bold math
\usepackage{physics}
\usepackage{ragged2e}
\usepackage{subcaption}
\usepackage{graphicx}
\usepackage{float}
\usepackage{lmodern}
\usepackage{amsmath}
\usepackage{xcolor}
\usepackage{hyperref}
\makeatletter
\def\Hy@appendixstring{Appendix}
\makeatother
\begin{document}

\preprint{APS/123-QED}

\title{All-optical switching in a trapped ion cavity QED system: a comparative study}% Force line breaks with \\

\author{Abhijit Kundu}
\affiliation{Department of Physics, Indian Institute of Technology Tirupati, Yerpedu-517619, Andhra Pradesh, India.}

\author{Vijay Bhatt}
\affiliation{Department of Physics, Indian Institute of Technology Tirupati, Yerpedu-517619, Andhra Pradesh, India.}

\author{Arijit Sharma}\email{arijit@iittp.ac.in}
\affiliation{Department of Physics, Indian Institute of Technology Tirupati, Yerpedu-517619, Andhra Pradesh, India.}
\affiliation{Center for Atomic, Molecular, and Optical Sciences and Technologies,
Indian Institute of Technology Tirupati, Yerpedu-517619, Andhra Pradesh, India.}

\date{\today}% It is always \today, today,

% \begin{abstract}
% We have investigated cavity-EIT based all optical switching theoretically in a time dependent frame work. It was observed that the time dependent behavior gives more insights than the steady state behavior of switching. Three different switching schemes are analyzed and compared to metricize the switching performance with respect to relevant important controlling parameters.  
% \end{abstract}

\begin{abstract}
We investigate the transient dynamics of cavity-EIT-based all-optical switching in a system of trapped ions coupled to an optical cavity through numerical simulations. In contrast to steady-state analysis, the time-dependent response provides direct insight into the switching speed, transient dynamics, and achievable switching contrast. We consider three distinct switching schemes and systematically compare their performance as a function of the relevant system parameters. The switching contrast is evaluated from the time-dependent cavity output and used to characterize the performance of each scheme. For the four-level N-type system, we find distinct trade-offs between switching through suppression of the cavity-EIT signal and switching through resonance shifting, with the former exhibiting a faster temporal response. The corresponding three-level scheme can achieve near-unity switching contrast with substantially shorter switching pulses, although it does not provide an independent switching control field. These results establish the performance limits and trade-offs of different cavity-EIT-based switching schemes and provide guidelines for optimizing their operation for applications such as high-speed optical gating, frequency-selective photon routing, frequency-multiplexed quantum communication, etc.
\end{abstract}
\maketitle

%\tableofcontents
\section{Introduction}
The ability to coherently manipulate light with light is a fundamental goal of atomic, molecular, and optical (AMO) physics, key to the realization of optical information processing, quantum communication and photonic quantum technologies. All‑optical switching, in which the propagation of a weak optical signal is modulated by another optical field without converting the signal into the electronic domain, is a key enabling mechanism for optical communication networks, quantum information processing, photonic logic, and low‑power optical computing \cite{Miller2010,Soljacic2004}. In such a switch, the control field modifies the optical response of the medium, enabling the probe field to be switched between distinct output states. Achieving such switching at low optical powers while maintaining high contrast, fast response, and minimal loss remains an important challenge because conventional nonlinear optical processes generally require either large optical intensities or long interaction lengths \cite{wang2022field,tseng2022efficient,thomas2019raman,srivathsan2025toward}.
\\

% Electromagnetically induced transparency (EIT) has become one of the most promising physical mechanisms for all-optical switching due to its ability to significantly alter the optical response of an atomic medium through quantum interference. Since its initial experimental observation, EIT has enabled numerous applications, such as slow and stopped light \cite{hau1999light,kash1999ultraslow,phillips2001storage}, quantum memories \cite{chaneliere2005storage,simon2007single,zhao2009long,EITQM2022}, enhanced nonlinear interactions \cite{lukin2001controlling,popov2005nonlinear,schmidt1996giant,harris1999nonlinear}, precision spectroscopy \cite{Marangos1998}, and low-light-level optical switching \cite{Boller1991,Harris1997,Fleischhauer2005,Lukin2003}. 
Atomic systems provide an extremely effective platform to achieve low power optical nonlinearities owing to the capacity of their narrow optical transitions to allow for strong light-matter interactions at quite low light fields. Among the various approaches, electromagnetically induced transparency (EIT) has become one of the most promising physical mechanisms for all-optical switching due to its ability to significantly alter the optical response of an atomic medium. Since its initial experimental observation\cite{Boller1991,Harris1997,Fleischhauer2005,Lukin2003}, EIT has been exploited for different applications, such as quantum interference \cite{hau1999light,kash1999ultraslow,phillips2001storage}, quantum memories \cite{chaneliere2005storage,simon2007single,zhao2009long,EITQM2022}, enhanced nonlinear interactions \cite{lukin2001controlling,popov2005nonlinear,schmidt1996giant,harris1999nonlinear}, precision spectroscopy \cite{Marangos1998}, and low-light-level optical switching \cite{PhysRevLett.81.3611,doi:10.1126/science.1110151,Zhang:07,Lee:12}.\\

The integration of EIT and cavity quantum electrodynamics (cavity-QED) offers an appealing platform for enhancing light-matter interactions through the strong coupling between atoms and a well-defined cavity mode. Cavity-EIT is in many respects more convenient than free-space EIT because it allows for an enhanced effective optical depth for narrow transmission resonances of the cavity and thus for an increased cooperativity. This allows for a very efficient manipulation of optical fields, even at the few-photon level. The unique properties of EIT in cavity configuration lead to a number of applications, such as efficient quantum memories \cite{lukin2000entanglement,dantan2004quantum,dantan2006dynamics,gorshkov2007photon}, Fock-state quantum filters \cite{nikoghosyan2010photon}, sensitive atomic magnetometers \cite{budker1999nonlinear,scully1992high}, cavity optomechanical cooling \cite{tan2011cooling,genes2011atom}, lasing without inversion \cite{wu2008evidence}, and all-optical switching \cite{nielsen2011efficient,tanji2011vacuum}  on very low optical power.
\\

Past research has shown the ability of all-optical switching in cavity-EIT systems, mostly focused on the steady-state optical response. Nevertheless, these investigations are important for the determination of operational regimes and possible switching contrasts, however, they do not fully characterize the transient behavior of the switch \cite{PhysRevA.85.013840,albert2011cavity}. It is equally important in both the cases of practical optical applications and quantum information processing since the switching dynamics define the switching rate, rise and fall times, and trade-off between the switching speed and contrast. A systematic characterization of the transient switching dynamics is therefore essential for determining the temporal performance of cavity-EIT-based optical switches.
\\

% There are some experimental evidences that shows switching with respect to time in N-type cavity-EIT based switching, but they reported the switching rate is limited by AOM's rise and fall time. Thus the actuall theoretical limit of switching rate with high contrast is an interesting question to ask. 
Experimental studies have also demonstrated time-dependent switching in N-type cavity-EIT systems. However, in these experiments, the observed temporal response was limited by the rise and fall times of the acousto-optic modulator (AOM) used to control the switching field \cite{Sheng:11, DUAN201673}. Consequently, the experimentally observed switching rate does not necessarily represent the intrinsic temporal limit imposed by the cavity-EIT system itself. This raises an important question: what is the fundamental switching timescale that can be achieved by the cavity-EIT system while maintaining a high switching contrast. Determining this intrinsic temporal limit is particularly important for assessing the suitability of cavity-EIT-based switching for high-speed optical and quantum-information applications.
\\

Here, we numerically investigate the transient dynamics of cavity-EIT-based all-optical switching in a system of trapped ions coupled to an optical cavity. We characterize the switching performance in terms of the switching contrast and systematically study its dependence on the relevant system parameters. In particular, we identify three distinct switching schemes based on different mechanisms for controlling the cavity-EIT response. The switching performance of these schemes is analyzed and compared in the subsequent sections, with particular emphasis on their achievable contrast, switching speed, and temporal response.
\\

% The all-optical switching mechanisms that are discussed here are useful for applications in quantum computing and quantum communication. The simple on-off of the output mode of the cavity could potentially applied to quantum gate operation. Switching where the frequency mode of the photon is selectively chosen by the switching field can be applied for frequency selective routing of photons to different quantum nodes operationg in different frequencies.

The all-optical switching mechanisms discussed here have potential applications in quantum computing and quantum communication. The on-off control of the cavity output mode could, for example, provide a mechanism for optical control in quantum gate operations. In addition, schemes in which the frequency mode of the emitted photon is selectively controlled by the switching field could enable frequency-selective routing of photons between quantum nodes operating at different frequencies.
\\

The remainder of this paper is organized as follows. In Sec.~II, we present the theoretical model of the cavity-QED system, including the Hamiltonian and master equation used in our analysis. In Sec.~III, we introduce and discuss the different switching schemes based on the cavity-EIT system. In Sec.~IV, we investigate the time-dependent behavior of the switching schemes, focusing on the transient dynamics of the cavity field and its response following the application and removal of the switching field. In Sec.~V, we quantify the switching performance in terms of the switching contrast and examine its dependence on the relevant switching-field parameters.

% On the other hand, cavity-EIT also leads to significantly enhanced cross-Kerr nonlinearities and thus supports a variety of nonlinear photonics applications, including a strong photon blockade \cite{imamoglu1997strongly,grangier1998comment,gheri1999quantum} and interactions of photons \cite{werner1999photon}, as well as generation of highly entangled states \cite{dantan2006spin} and even a quantum phase transition of light \cite{hartmann2006strongly}.

% Cavity-EIT has been experimentally demonstrated with a variety of media, including atomic beams \cite{muller1997optical}, cold and thermal atomic media \cite{hernandez2007vacuum,wu2008observation,laupretre2011photon}, single atom trapped inside high-finesse optical cavities \cite{mucke2010electromagnetically,kampschulte2010optical}, and cold ion Coulomb crystals \cite{albert2011cavity}.

\section{THEORETICAL MODEL}\label{sec:roadmap}
% The system consists of multi-trapped ${}^{40}\mathrm{Ca}^{+}$ ions strongly coupled to a single-mode high-finesse optical cavity. 
We consider a four-level N-type atomic system comprising two long-lived ground (metastable) states and two excited states. The cavity-EIT is established between two ground states and one excited state in a lambda type structure\cite{albert2011cavity,mucke2010electromagnetically}. The switching field is applied from one ground state to another excited state. This forms a typical four-level EIT system. This type of system has been studied for large Kerr nonlinearity \cite{PhysRevLett.91.093601}, optical switching \cite{PhysRevA.68.041801,PhysRevLett.102.203902}, quantum interference\cite{PhysRevLett.81.3611} etc.   
\\

\begin{figure}[htbp]
    
    \includegraphics[width=\columnwidth]{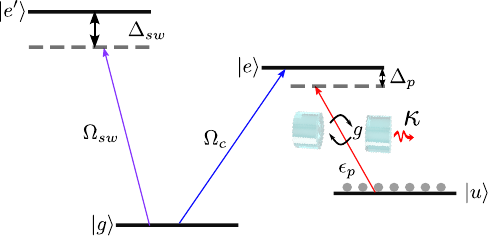}
    \caption{\justifying{
    Switching scheme for a four-level N-type system. The cavity-EIT is established between level $\ket{g}$, $\ket{e}$ and $\ket{u}$. A weak probe of amplitude $\epsilon_p$ is applied between  $\ket{u}$ and $\ket{e}$. The switching field is applied between the ground state, $\ket{g}$, and the excited state, $\ket{e'}$, with Rabi frequency $\Omega_{sw}$ for different detunings of $\Delta_{sw}$. 
    }}
    \label{fig_sw_scheme}
\end{figure}

The presence of the cavity enhances the interaction between the atoms and the cavity. Additionally cavity makes the atoms to interact with well defined spatiotemporal field modes which provides great level of control. Here we took the ${}^{40}\mathrm{Ca}^{+}$ ion cloud as reference atomic medium and the parameters related to atom are chosen according to that. The four relevant energy levels along with the cavity is illustrated in \autoref{fig_sw_scheme}. The $\ket{u}\leftrightarrow\ket{e}$ transition is coupled to a single cavity mode with atom-cavity coupling strength $g$, while the cavity is weakly driven by a probe field of amplitude $\eta$. A strong classical control field with Rabi frequency $\Omega_c$ resonantly drives the $\ket{g}\leftrightarrow\ket{e}$ transition, giving rise to cavity electromagnetically induced transparency (cavity-EIT) through destructive quantum interference.
\\

The interaction of the four-level system with the quantized cavity field and the classical driving fields is described under the electric-dipole and rotating-wave approximations. By transforming into a frame rotating at the frequencies of the applied fields and neglecting the rapidly oscillating terms, the effective Hamiltonian of the coupled atom-cavity system can be written as

\begin{equation}
\begin{aligned}
H=&H_0+H_I,
\\
H_0=&-\Delta_p a^\dagger a
-\Delta_p\sigma_{ee}-\Delta_p\sigma_{gg}
-(\Delta_{sw}+\Delta_p)\sigma_{e'e'},
\\
H_I=&\frac{g}{\sqrt{2}}(a^\dagger\sigma_{ue}
+\sigma_{eu} a)
+\eta(a+a^\dagger)
\\
&
+\frac{\Omega_c}{\sqrt{2}}
(\sigma_{ge}+\sigma_{eg})
+\frac{\Omega_{sw}}{\sqrt{2}}
(\sigma_{ge'}+\sigma_{e'g}).
\end{aligned}\label{eq:Ham}
\end{equation}

Here, $a$ ($a^{\dagger}$) denotes the annihilation (creation) operator of the cavity mode, while $\sigma_{ij}=|i\rangle\langle j|$ are the atomic transition operators. The parameter $g$ represents the atom-cavity coupling strength, $\eta$ is the amplitude of the weak probe field driving the cavity, and $\Omega_c$ and $\Omega_{sw}$ denote the Rabi frequencies of the control and switching fields, respectively.
The detunings are defined as
\begin{equation}
\Delta_p = \omega_p-\omega_{\mathrm{cav}}; 
\hspace{0.2cm} \Delta_c = \omega_c-\omega_{eg};
\hspace{0.2cm}
\Delta_{\mathrm{sw}}
= \omega_{\mathrm{sw}}-\omega_{e'g},
\label{eq:detunings}
\end{equation}

where $\omega_p$, $\omega_c$, and $\omega_{\mathrm{sw}}$ denote
the frequencies of the probe, control, and switching fields,
respectively, $\omega_{\mathrm{cav}}$ is the resonance frequency
of the cavity mode, and $\omega_{eg}$ and $\omega_{e'g}$ are the
transition frequencies of the $|g\rangle\leftrightarrow|e\rangle$
and $|g\rangle\leftrightarrow|e'\rangle$ transitions, respectively. The factor $1/\sqrt{2}$ comes in cavity coupling and Rabi frequencies because the ion trapped is considered to be in thermal motion which is the usual case for a trapped ion system and the motion is called the secular frequency. The secular motion is faster than the dynamics of cavity-EIT so the effect is average out and gives a scaling factor $1/\sqrt{2}$ for the coupling parameters \cite{PhysRevA.85.013840}.

 \section{Switching scheme}
Typically, optical switching directs photons between two spatial output ports upon application of a control field, enabling high-speed optical routing in communication and photonic networks \cite{r7xf-lptn,10.1063/1.5115814}. In contrast, another important class of optical switches operates in the frequency domain, where the control field determines in which spectral mode the photon is emitted without altering the propagation path. Such frequency-selective switching is particularly advantageous for quantum networks, where photons of different frequencies can interface with heterogeneous quantum systems, enable frequency-multiplexed quantum communication \cite{PhysRevLett.113.053603}, and enable spectrally selective routing of photons between different quantum nodes \cite{PhysRevLett.109.147404}.
\\
\begin{figure}[htbp]
    \includegraphics[width=\columnwidth]{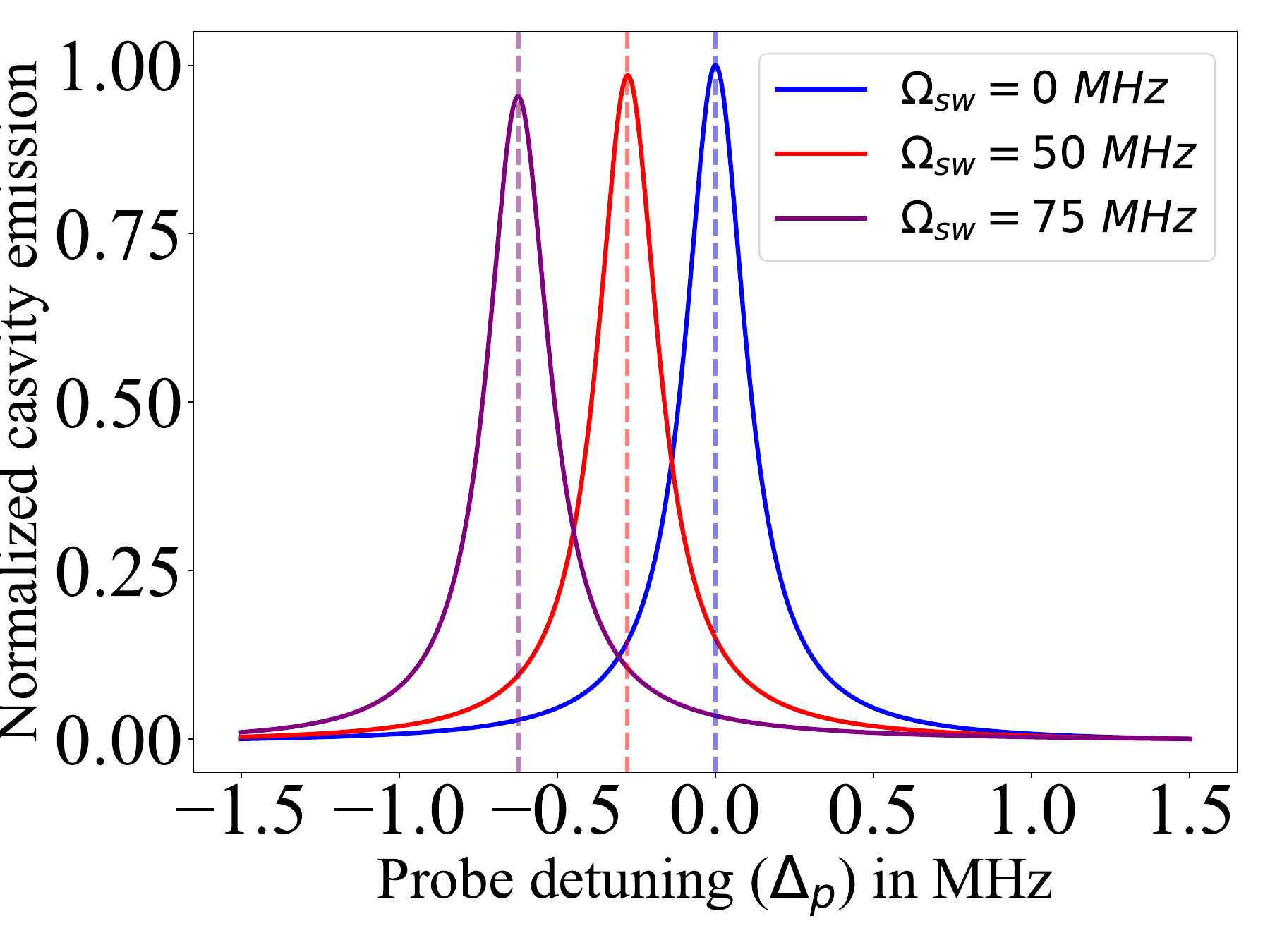}
    \caption{\justifying{
    Cavity-EIT-based optical switching with shifting of the cavity-EIT peak. The parameters are taken as $\kappa_1=1.53$ MHz, $\kappa_2=7.85$ kHz,           $\kappa_A=0.67$ MHz, $\gamma_{eg}=11.7$ MHz, $\gamma_{e'g}=11.6$ MHz, $g_0=0.54$ MHz, $N=1000$, $\Omega_c=4.35$ MHz, $\Omega_{sw}=75$ MHz.
    }}
    \label{fig_sw_shifting}
\end{figure}

% In our N-type system, such switching in the frequency domain can be readily realized. When the detuning of the switching laser from the excited state, $\ket{e'}$ is much larger than the linewidth of the excited state ($\Delta_{sw}\gg \Gamma_{e'}$), there will be barely any transition to the excited state, $\ket{e'}$ rather the ground state will be Stark shifted by an amount $\Omega_{sw}^2/(2\Delta_{sw})$. Due to this induced shift, the two-photon resonance condition of the EIT changes, and the central frequency of the cavity-EIT spectrum also shifts by the same amount. If the switching field is strong enough that the Stark shift is larger than the linewidth of the cavity-EIT spectrum, a significant shift is observed in the spectrum, and it can be distinguished as a separate mode.  

In our N-type system, switching can instead be realized in the frequency domain. When the switching laser is detuned far from the excited state, $\ket{e'}$, such that $\Delta_{sw} \gg \Gamma_{e'}$, excitation of $\ket{e'}$ is strongly suppressed. Instead, the off-resonant interaction induces an AC Stark shift of the ground state $\ket{g}$, given by $\Omega_{sw}^2/(2\Delta_{sw})$. Consequently, the two-photon resonance condition for cavity-EIT is modified, resulting in an identical shift of the central frequency of the cavity-EIT resonance.
\\

If the switching field is sufficiently strong that the induced Stark shift exceeds the linewidth of the cavity-EIT resonance, the shifted resonance becomes spectrally resolved from the one without applying the switching field. Thus, by applying the switching field, the cavity output can be channelized into a different mode. This type of switching has previously been demonstrated  experimentally with the reflected signal of the cavity \cite{albert2011cavity}. \\

We perform the simulation considering the Hamiltonian defined in \autoref{eq:Ham} using the master equation approach. The parameters are chosen the same as taken in ref. \cite{albert2011cavity}. In \autoref{fig_sw_shifting}, we have shown the shifting of the cavity resonance, in the transmitted output of the cavity for three different values of switching field intensity ($\Omega_{sw}$). This shows that based on the switching field intensity, the cavity mode can be switched to many different modes.
\\

\begin{figure}[htbp]
\begin{subfigure}[t]{\linewidth}
\includegraphics[width=\linewidth]{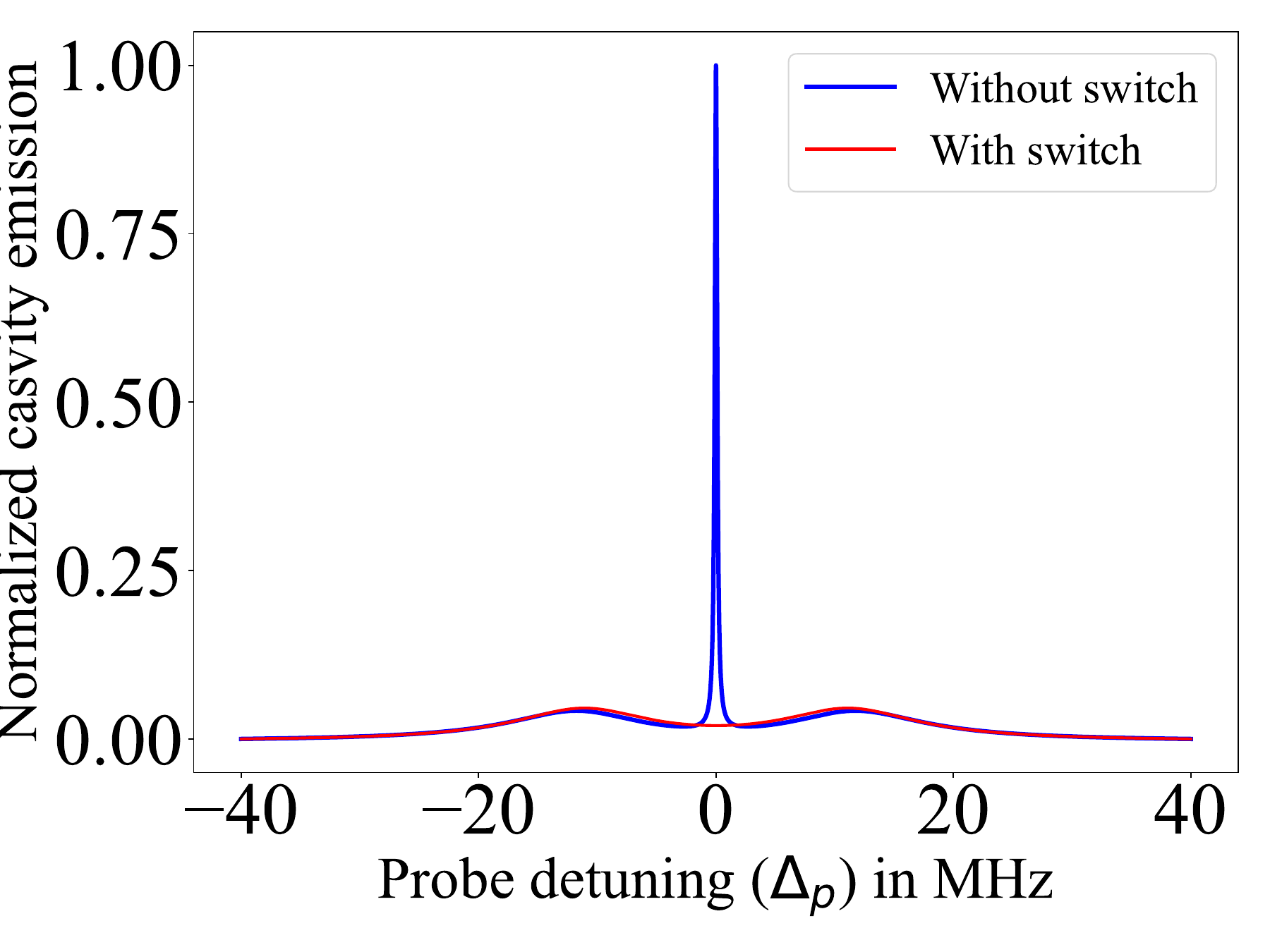}
\caption{}
\label{fig:sw_suppress_a}
\end{subfigure}
\hfill
\begin{subfigure}[t]{\linewidth}
\includegraphics[width=\linewidth]{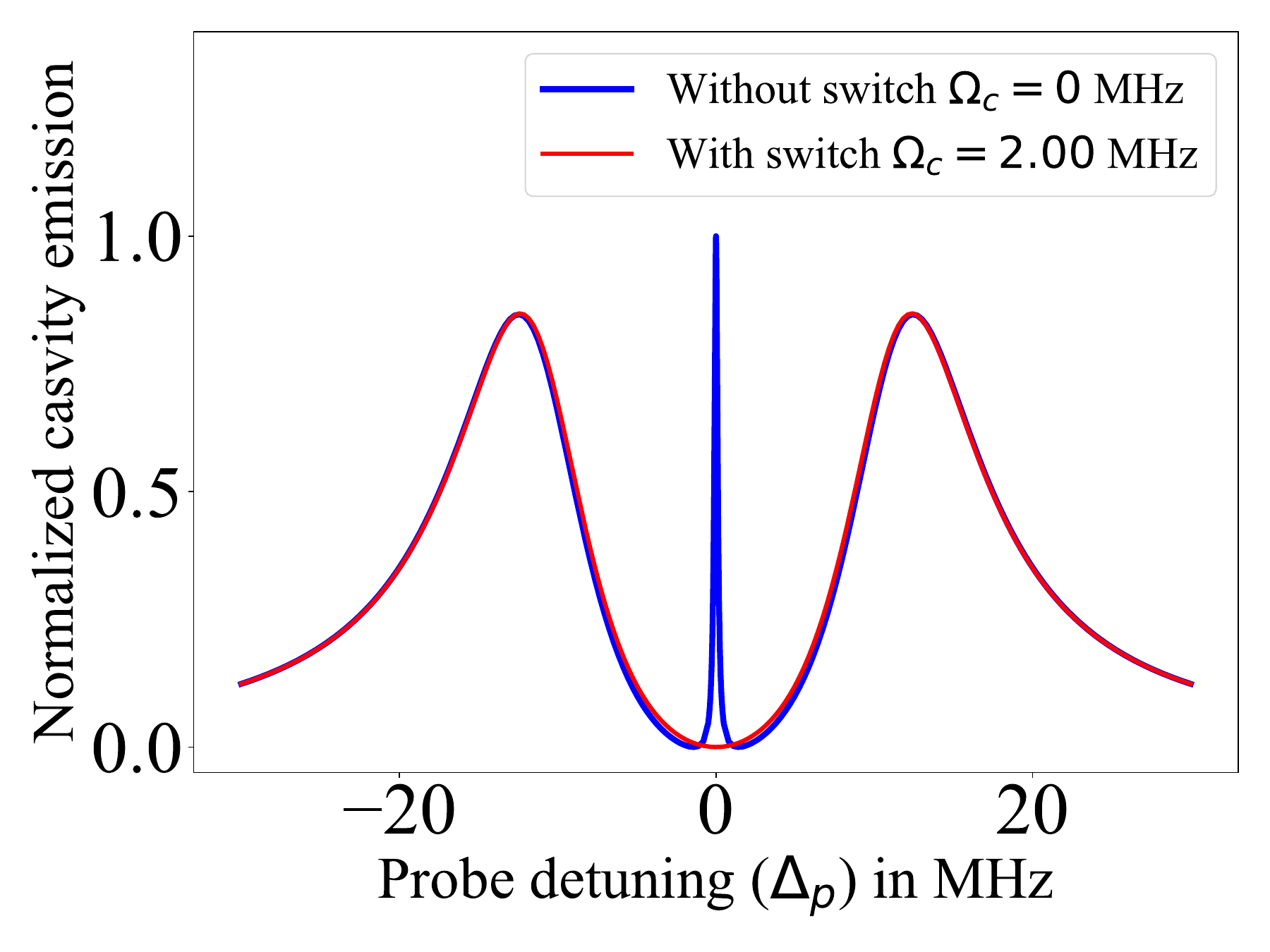}
\caption{}
\label{fig:sw_suppress_b}
\end{subfigure}
\caption{\justifying {Cavity-EIT-based optical switching with suppression of cavity-EIT peak. (a) In the case of a 4-level system. The parameters are taken as $\kappa_1=1.53$ MHz, $\kappa_2=7.85$ kHz,$\kappa_A=0.67$ MHz, $\gamma_{eg}=11.7$ MHz, $\gamma_{e'g}=11.6$ MHz, $\gamma_{gu}=1$ kHz, $g_0=0.54$ MHz, $N=996$, $\Omega_c=4.35$ MHz, $\Omega_{sw}=75$ MHz. (b) In the case of a 3-level system, with parameters, $\kappa_1=1.53$ MHz, $\kappa_2=7.85$ kHz,$\kappa_A=0.67$ MHz, $\gamma_{eg}=11.7$ MHz, $\gamma_{eu}=0.84$ MHz, $\gamma_{gu}=1$ kHz, $g_0=0.54$ MHz, $N=1000$, $\Omega_c=2$ MHz.}
}
\label{fig:sw_supress}
\end{figure}

Another switching mechanism that can be implemented in our system is the periodic on-off modulation of the cavity-EIT signal by applying the switching field in a pulsed manner. Such controlled suppression of an optical signal is relevant for realizing optical transistors and high-fidelity optical gate operations \cite{PhysRevLett.113.053601,Tiarks2019}. When the switching field is resonant with the $\ket{g}\rightarrow\ket{e'}$ transition, the cavity-EIT signal is strongly suppressed. Under this condition, the switching field drives the population from the ground state $\ket{g}$ to the excited state $\ket{e'}$, thereby modifying the atomic response experienced by the cavity field. In the regime considered here, the linear susceptibility is strongly suppressed, while the nonlinear contribution to the susceptibility becomes dominant \cite{PhysRevLett.81.3611,Chen:05}. The resulting suppression of the cavity-EIT signal is illustrated in \autoref{fig:sw_suppress_a}, which shows the simulated cavity-EIT spectrum in the absence and presence of the switching field.
\\

\par
A similar on-off switching operation can also be realized using a conventional three-level cavity-EIT system by switching the control field itself. When the control field is turned off, the EIT condition is destroyed and the three-level system effectively reduces to a two-level atom coupled to the cavity. In the strong-coupling regime, the resulting cavity spectrum exhibits vacuum Rabi splitting rather than a cavity-EIT resonance \cite{PhysRevLett.93.233603,PhysRevLett.53.1732}. Thus, the cavity-EIT signal can be strongly suppressed by simply turning off the control field. Such control of cavity transmission through switching of the control field has also been experimentally demonstrated \cite{PhysRevA.96.033813}. For a direct comparison with the four-level scheme, we consider the three-level subsystem $\{\ket{g},\ket{e},\ket{u}\}$ of the four-level system shown in \autoref{fig_sw_scheme}, together with the cavity mode. The corresponding switching dynamics are illustrated in \autoref{fig:sw_suppress_b}.
\\

\par
The key distinction between the two schemes is the mechanism used to suppress the cavity-EIT signal. In the conventional three-level configuration, the same control field that establishes the EIT condition also serves as the switching field. Consequently, switching the output from the ON to the OFF state requires switching the EIT control field off, thereby destroying the dark-state condition. In contrast, the four-level N-type system provides an independent switching channel through the additional $\ket{g}\rightarrow\ket{e'}$ transition. The cavity-EIT control field can therefore remain continuously applied, while a separate resonant switching field controls the suppression of the cavity-EIT signal. This separation of the EIT and switching fields provides an additional degree of freedom for optimizing the operating point and the switching dynamics independently.
\\

\par
In the following section, we compare the time-domain dynamics of the two switching schemes. Such a comparison provides a direct assessment of their switching speed, extinction, and transient response, and allows us to identify the potential advantages and application regimes of each approach.

\section{Time dynamics of switching schemes}
The analysis of the steady state finds out the change in the cavity-EIT transmission because of the switching field, however it does not give direct information regarding the time response of the switching mechanism. In particular, the switching time is governed by the coupled dynamics of the intracavity field and the atomic coherences following the application or removal of the switching field. To investigate these transient dynamics, we solve the time-dependent master equation of the coupled atom-cavity system.

The density matrix $\rho$ of the atom-cavity system evolves according to the Lindblad master equation

\begin{equation}
\frac{d\rho}{dt}
=
-i[H(t),\rho]
+
\sum_j \mathcal{D}[L_j]\rho ,
\label{eq:master}
\end{equation}
where
\begin{equation}
\mathcal{D}[L_j]\rho
=
L_j\rho L_j^\dagger
-\frac{1}{2}
\left(
L_j^\dagger L_j\rho
+
\rho L_j^\dagger L_j
\right)
\label{eq:dissipator}
\end{equation}
is the Lindblad dissipator and $L_j$ denotes the corresponding decay operator. The Hamiltonian $H(t)$ is the atom-cavity Hamiltonian introduced in Sec.~II.

The cavity dissipative dynamics include leakage through the two cavity mirrors and intracavity absorption. The collapse operators describing these three channels are,

\begin{equation}
L_{\kappa_j}=\sqrt{\kappa_j}\,a,
\qquad
j=1,2,A,
\label{eq:cavity_collapse}
\end{equation}
where $\kappa_1$ and $\kappa_2$ denote the decay rates through the
two cavity mirrors, while $\kappa_A$ accounts for intracavity
absorption.

The spontaneous decay of the excited atomic states is included
through the collapse operators
\begin{equation}
L_{\gamma_{ij}}
=
\sqrt{\gamma_{ij}}\,\sigma_{ji},
\qquad
(ij)=(eg),(e'g),
\label{eq:atomic_collapse}
\end{equation}
where $\gamma_{eg}$ and $\gamma_{e'g}$ are the corresponding
spontaneous-emission rates.
\\

% \begin{figure}[htbp]
%     \includegraphics[width=\columnwidth]{switching_with_time.pdf}
%     \caption{\justifying{
    
%     }}
%     \label{fig_sw_suppress}
% \end{figure}

% The process of switching takes place by exposing the system to the switching field specifically set to the cavity-EIT state. 

The switching process occurs by periodically applying the switching field. Without the switching field, the input probe field will undergo the cavity-EIT transition due to the destructive interference between the relevant excitation pathways. After applying the switching field, the extra coupling modifies the atomic coherence and consequently alters the optical response of the cavity. As a result, the atomic state and the intracavity field evolve gradually towards the new steady state instead of changing instantaneously.
\\

\begin{figure}[htbp]
\begin{subfigure}[t]{\linewidth}
\includegraphics[width=\linewidth]{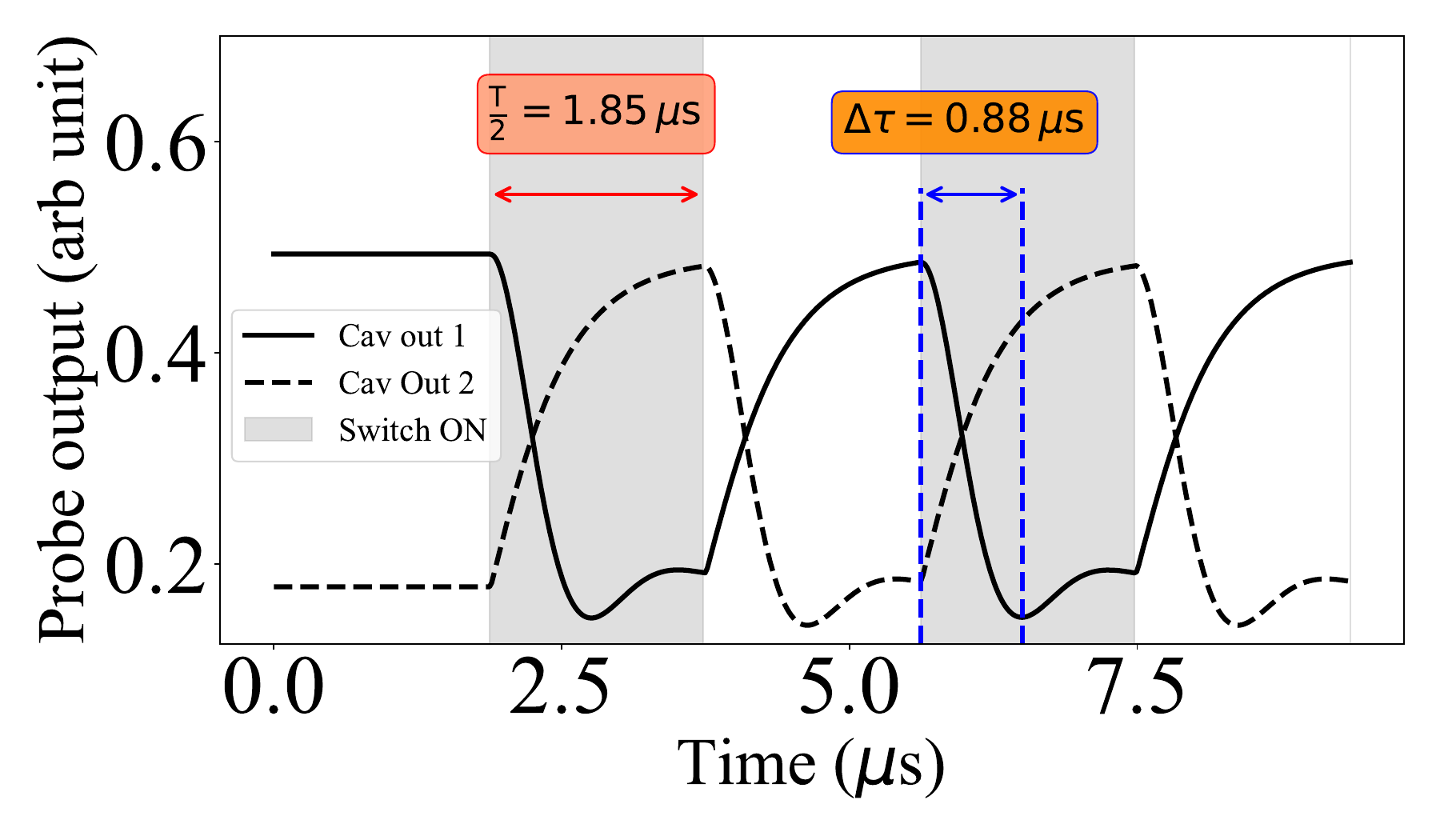}
\caption{}
\label{fig:sw_time_4lvl_a}
\end{subfigure}
\hfill
\begin{subfigure}[t]{\linewidth}
\includegraphics[width=\linewidth]{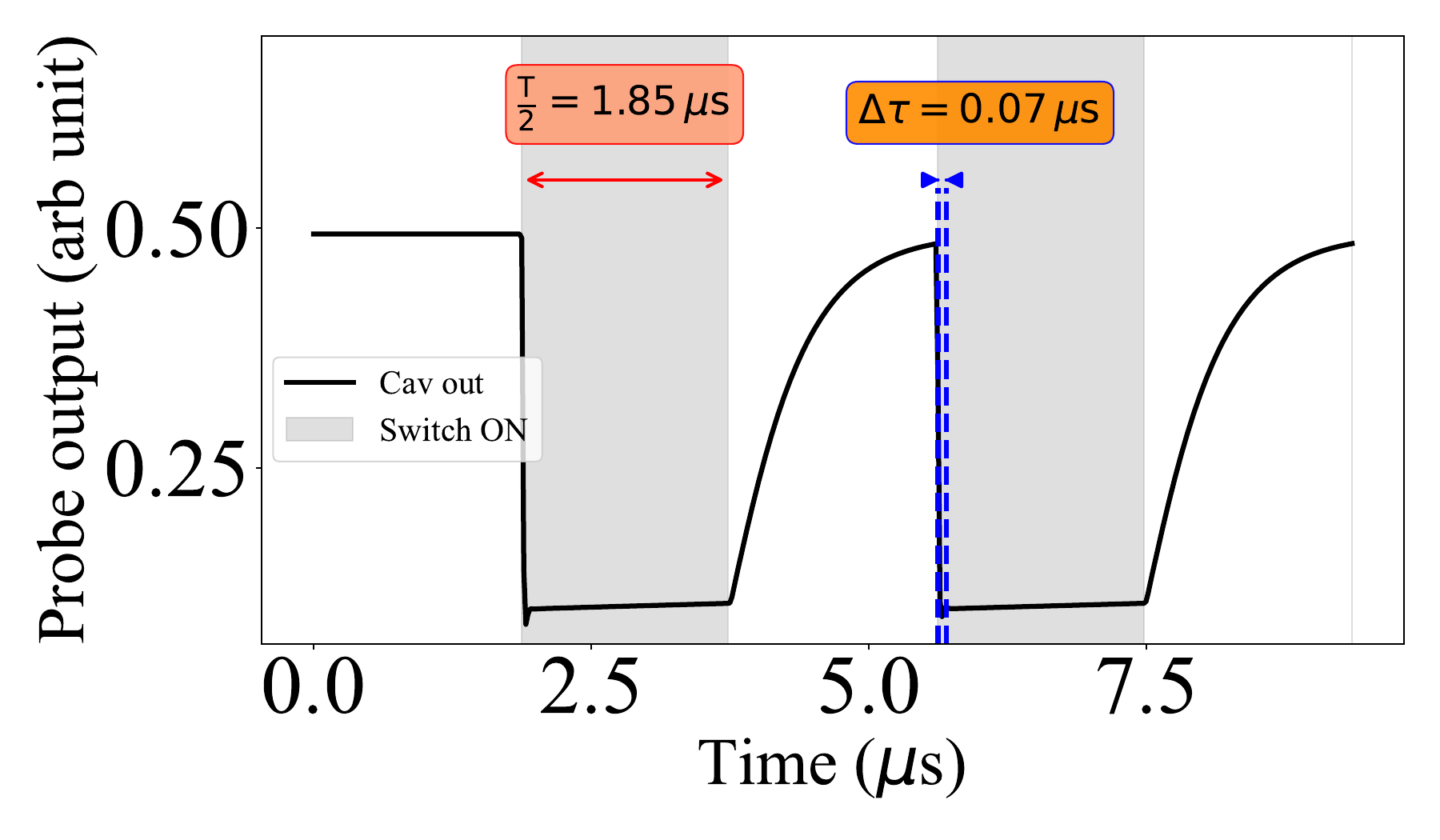}
\caption{}
\label{fig:w_time_4lvl_b}
\end{subfigure}
\caption{\justifying {Time dynamics of optical switching. (a) Switching with spectral shifting of the cavity transmission. (b) Switching with suppression of the cavity transmission. Except for the switching-laser detuning ($\Delta_{sw}$), which is $4.3$ GHz for case (a) and $0$ for case (b), respectively, all other parameters are identical for both cases and are given by $\kappa_1=1.53$ MHz, $\kappa_2=7.85$ kHz, $\kappa_A=0.67$ MHz, $\gamma_{eg}=11.7$ MHz, $\gamma_{e'g}=11.6$ MHz, $g_0=0.54$ MHz, $N=1000$, $\Omega_c=4.35$ MHz, and $\Omega_{sw}=75$ MHz.}
}
\label{fig:sw_time_4lvl}
\end{figure}

For a specified switching-field Rabi frequency ($\Omega_{\mathrm{sw}}$)  and detuning ($\Delta_{\mathrm{sw}}$), Eq.~\ref{eq:dissipator} is solved in time numerically, starting with the steady-state density matrix for the case of no switch field applied. The switching field is incorporated as a periodic function of square shape. To determine the temporal response of the switch, first, the dynamics of the cavity field are evaluated with time. Specifically, the emission of the cavity is computed using the expectation value of the cavity field operator,
\begin{equation}
I_{\mathrm{cav}}(t)
\propto
\left\langle a^\dagger a \right\rangle_t
=
\mathrm{Tr}
\left[
a^\dagger a\,\rho(t)
\right],
\label{eq:cavity_emission}
\end{equation}

\autoref{fig:sw_time_4lvl} depicts the temporal response of the cavity emission for two different schemes of 4-level cavity-EIT-based switching.
 The system is still in the initial cavity-EIT state before we apply the switching field. The cavity emission is roughly constant. After application of the switching field the atomic coherences evolve and the cavity emission changes accordingly. The system after the transient evolution tends to the steady state corresponding to the switched configuration. Thus the characteristic time ($\Delta \tau$) for the cavity emission to evolve from its initial value to the switched value directly measures the switching speed.
\\

% In case of switching by shifting of resonance the two modes becomes periodically switched on and off in opposite manner. In case of switching by suppression the cavity output is inhibited and restored periodically. After the application of switching field the cavity output jumps to minima very quickly. This action is much faster for switching by suppression compared to the switching by shifting with same set of parameters. After turning the switching field off, the system needs to establish the steady state again, thus the switching time is constrained by the cavity build-up time.   

In the case of switching through resonance shifting, the two spectral modes are periodically switched on and off in a complementary manner. That is, when one mode is switched on, the other is simultaneously switched off, resulting in an effective transfer of the cavity emission between the two modes. 
\\

In contrast, for switching through suppression, the cavity-EIT output is periodically inhibited and restored by applying and removing the switching field. When the switching field is applied, the cavity output rapidly drops to its minimum value. For the same set of system parameters, this suppression occurs significantly faster than the corresponding switching process based on resonance shifting. This difference arises because suppression does not require the cavity-EIT resonance to adiabatically shift from one spectral position to another; instead, the switching field directly inhibits the cavity-EIT output. 
\\

For both the cases, when the switching field is subsequently turned off, the cavity-EIT signal does not recover instantaneously. The system must re-establish the steady-state EIT condition, while the cavity field builds up toward its steady-state value. Consequently, the switching rate is ultimately constrained by the cavity buildup time, which is determined by the effective decay rate of the cavity-EIT mode. In the EIT regime, this effective decay rate is given by
\begin{equation}
    \kappa_{\mathrm{EIT}}
    =
    \gamma_{gu}
    +
    \kappa
    \frac{\Omega_c^2/2}{g_0^2N},
\end{equation}
where $\gamma_{gu}$ is the ground-state coherence decay rate, $\kappa$ is the bare cavity decay rate, $\Omega_c$ is the control-field Rabi frequency, $g_0$ is the single-atom cavity coupling strength, and $N$ is the number of atoms.
\\

The characteristic timescale for the recovery of the cavity-EIT signal is therefore determined by $\kappa_{\mathrm{EIT}}$. In particular, the switching field needs to be applied with a time period comparable to the characteristic response time ($T_{switch}=2\tau_{EIT}$), where, 
\begin{equation}
    \tau_{\mathrm{EIT}} \sim \frac{1}{2\kappa_{\mathrm{EIT}}},
\end{equation}
 % Thus, the minimum pulse duration required for complete switching is ultimately governed by the cavity-EIT response time. In fact the characteristic timescale can also be predicted directly from the linewidth of the steady-state cavity-EIT spectrum. But the steady state behavior lacks in explaining how fast it drops from high to low, and what the maximum switching rate is that can be achieved without compromising the contrast. These are important to know in case of optical switching to perform the gate operations or optical transistor in action.
 Thus, the minimum pulse duration required to achieve complete switching is ultimately governed by the cavity-EIT response time. The characteristic timescale can, in principle, be estimated directly from the linewidth of the steady-state cavity-EIT spectrum. However, a steady-state spectral analysis alone does not reveal the transient response of the system, particularly how rapidly the cavity output falls from the high-transmission state to the suppressed state following the application of the switching field. It also does not determine the maximum switching rate that can be achieved while maintaining a sufficiently high switching contrast. These transient characteristics are crucial for assessing the performance of the switching scheme in practical applications, such as high-fidelity optical gate operations and optical transistor action, where both the switching speed and the achievable contrast are essential.

% We also perform similar time-domain analysis for 3-level switching. In this case the control field itself is turned off, destroying the EIT itself. Unlike previous case, here the switching rate does not restricted by the cavity buildup time. It is much faster than the characteristic time of cavity-EIT.

\begin{figure}[htbp]
    \includegraphics[width=\columnwidth]{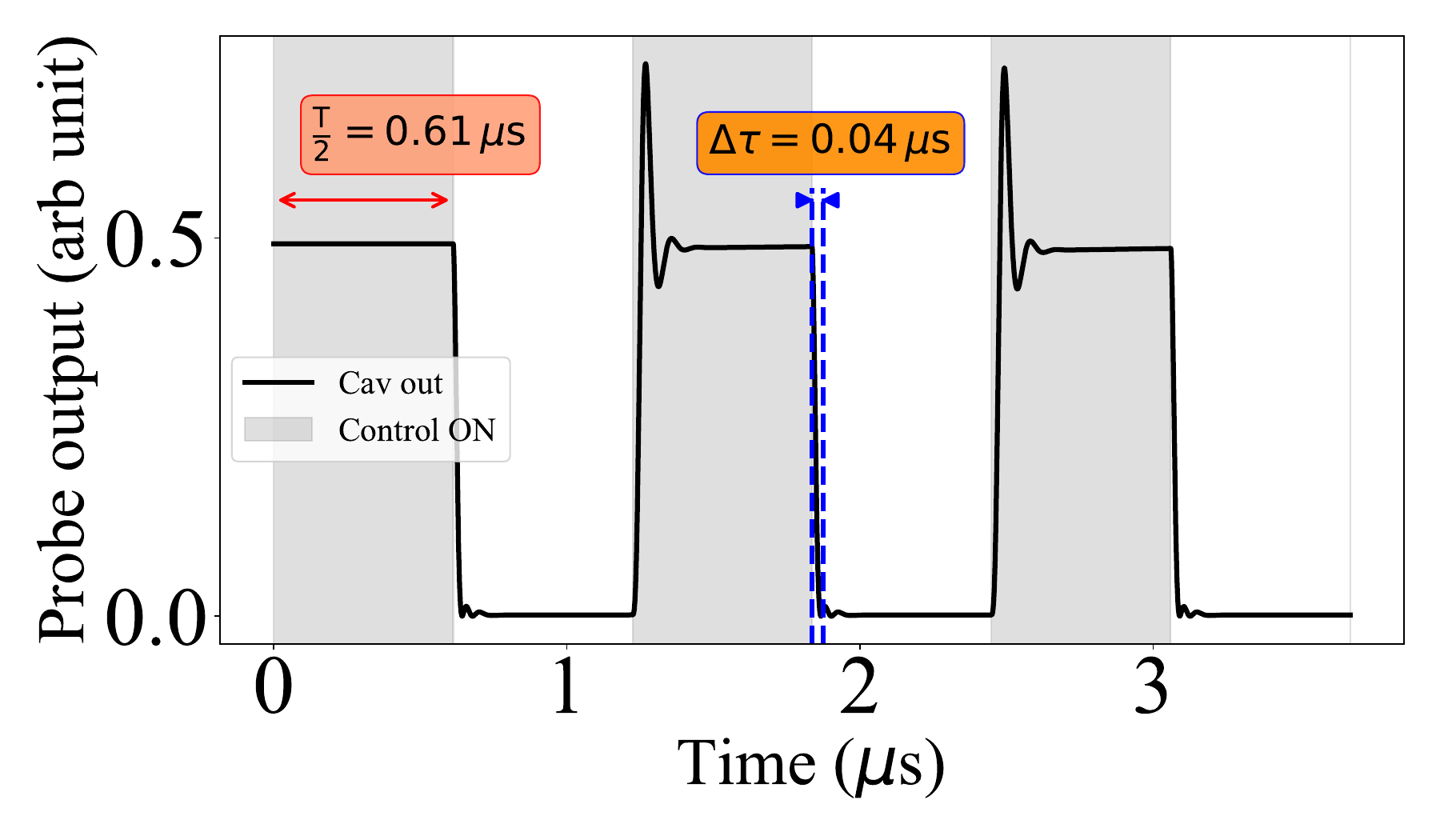}
    \caption{\justifying{ Time dynamics of optical switching for three-level cavity-EIT based switching. The parameters are taken as $\kappa_1=1.53$ MHz, $\kappa_2=7.85$ kHz, $\kappa_A=0.67$ MHz, $\gamma_{eg}=11.7$ MHz, $\gamma_{eu}=0.84$ MHz, $\gamma_{gu}=1$ kHz, $g_0=0.54$ MHz, $N=1000$, $\Omega_c=2$ MHz.
    }}
    \label{fig: sw_time_3level}
\end{figure}

We also perform a similar time-domain analysis for the three-level switching scheme. In this case, the control field itself is switched off, thereby directly destroying the cavity-EIT condition. Consequently, the suppression of the cavity-EIT signal is not limited by the cavity-EIT buildup time. Once the control field is removed, the EIT dark-state condition is destroyed directly, allowing the cavity output to drop on a timescale much shorter than the characteristic response time of the cavity-EIT resonance (see \autoref{fig: sw_time_3level}). Thus, in contrast to the four-level switching scheme, the switching rate of the three-level system can substantially be faster than the time extracted from the inverse of the cavity-EIT linewidth. This difference highlights the distinct dynamical mechanisms underlying the two switching schemes and motivates a direct comparison of their transient response and achievable switching contrast.

% \begin{figure}[htbp]
    
%     \includegraphics[width=\columnwidth]{switching_with_time_supp.pdf}
%     \caption{\justifying{
%     The parameters are taken as $\kappa_1=1.53$ MHz, $\kappa_2=7.85$ kHz,$\kappa_A=0.67$ MHz, $\gamma_{eg}=11.7$ MHz, $\gamma_{e'g}=11.6$ MHz, $g_0=0.54$ MHz, $N=996$, $\Omega_c=4.35$ MHz, $\omega_{sw}=75$ MHz.
%     }}
%     \label{fig_sw_suppress}
% \end{figure}

\section{Analysis of switching performance}

We characterise the switching performance in terms of both switching contrast and switching time. The switching contrast is then given by

\begin{equation}
C =
\frac{I_{\mathrm{on}}-I_{\mathrm{off}}}
     {I_{\mathrm{on}}+I_{\mathrm{off}}},
\label{eq:contrast}
\end{equation}

where $I_{\mathrm{on}}$ and $I_{\mathrm{off}}$ represent the cavity-emission
levels in the associated switched and unswitched conditions. The time of switching
is calculated from the minimum time for which the switching field is to be applied to get the switching contrast more than 0.5.

% Next, we systematically study the effect of switching-field parameters on the transient switching response, particularly its Rabi frequency
% $\Omega_{\mathrm{sw}}$ and detuning $\Delta_{\mathrm{sw}}$. This analysis helps us understand how the strength and the place in the spectrum of the switching field affect how fast the cavity-EIT response changes. Along, with figuring out the switching contrast the time-dependent calculation also gives the switching time and shows the balance between how fast the switch works and how clear the contrast is. 
\begin{figure}[htbp]
    
    \includegraphics[width=\columnwidth]{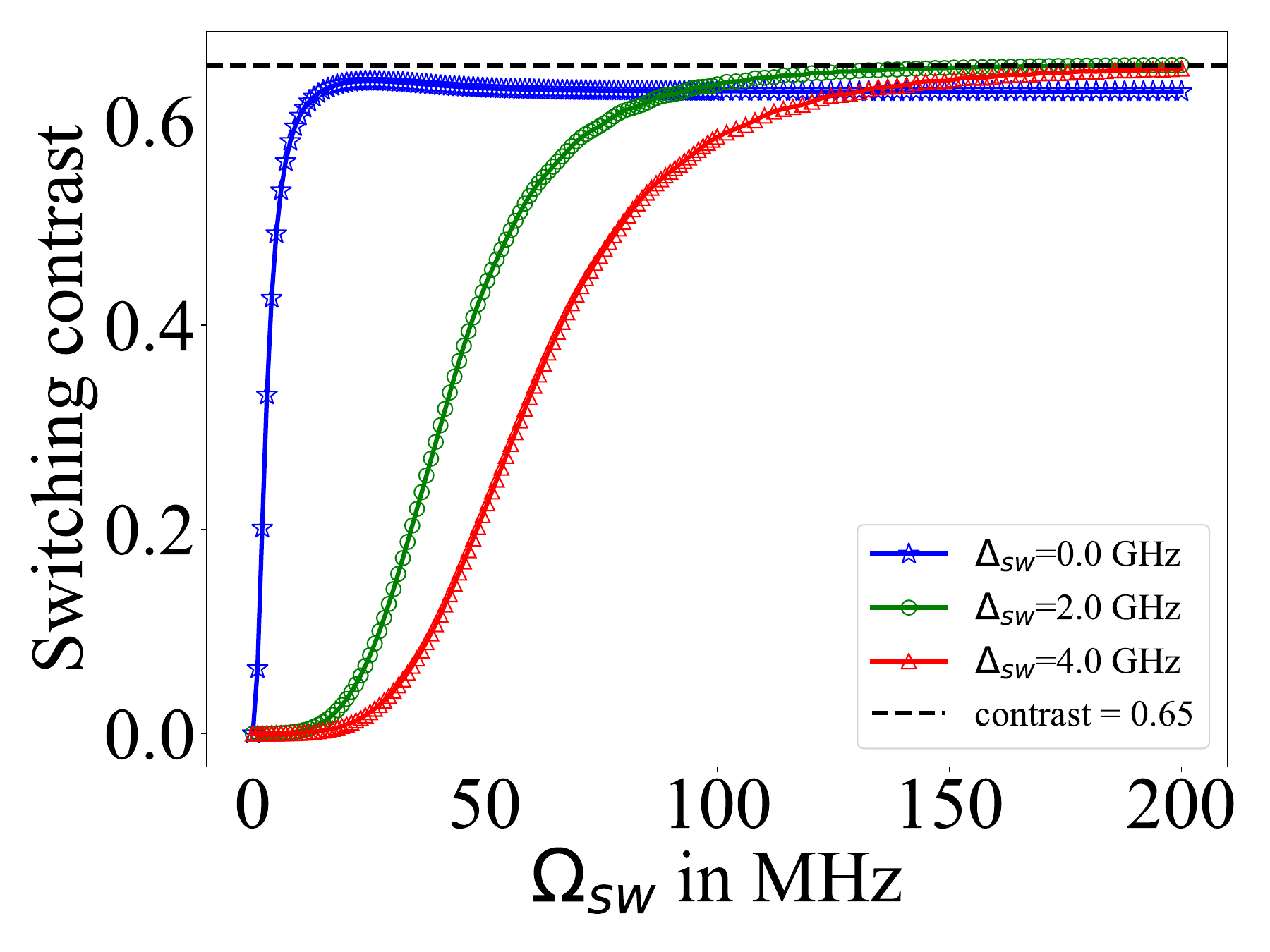}
    \caption{\justifying{ Switching contrast as a function of switching laser Rabi frequency ($\Omega_{sw}$) for 4-level switching. The parameters are taken as $\kappa_1=1.53$ MHz, $\kappa_2=7.85$ kHz,$\kappa_A=0.67$ MHz, $\gamma_{eg}=11.7$ MHz, $\gamma_{e'g}=11.6$ MHz, $g_0=0.54$ MHz, $N=1000$, $\Omega_c=4.35$ MHz, $\Omega_{sw}=75$ MHz.
    }}
    \label{fig:contrast_vs_Omega_sw_4lv}
\end{figure}

% The intensity of the switching field is a critical parameter that determines how much the cavity-EIT signal will be suppressed or how efficiently it is transferred to other modes. Consequently, the switching contrast depends on the intensity of the switching field. We analyzed the switching contrast with respect to the Rabi frequency of the switching in 4 level N-type system for both the cases of suppression and shifting. The results are illustrated in \autoref{fig:contrast_vs_Omega_sw_4lv}. Where switching in case of suppression reaches its maximum value of switching contrast at low switching field intensity, the switching in case of shifting requires more power to reach its maximum contrast. In case of only suppression, after reaching the maximum, further increasing the switching field intensity actually diminishes the switching contrast. It is because as the power increases the atoms undergo light induced shift, which tends to destroys the EIT coherence, as a result the empty cavity behavior tends to appear which is diminishing the switching contrast.     
The intensity of the switching field is a critical parameter that determines the extent to which the cavity-EIT signal is suppressed or the resonance is shifted. Consequently, the switching contrast is strongly dependent on the Rabi frequency of the switching field, $\Omega_{sw}$. We investigate this dependence for the four-level N-type system for both switching mechanisms, namely, suppression and resonance shifting, as shown in \autoref{fig:contrast_vs_Omega_sw_4lv}. For the suppression scheme, the switching contrast reaches its maximum value at a relatively low switching-field intensity. Beyond this optimum value, further increasing the switching-field intensity leads to a reduction in the switching contrast. In contrast, the resonance-shifting scheme requires a substantially higher switching-field intensity to achieve its maximum contrast, since the induced spectral shift must become sufficiently large compared with the cavity-EIT linewidth for the two modes to be well resolved.
\\

The reduction in contrast at high switching-field intensities in the suppression scheme can be attributed to the increasing influence of the strong switching field on the atomic coherence. As the switching-field intensity is increased, the resonant interaction induces significant light-induced shifts and additional broadening of the atomic levels, thereby perturbing the coherence responsible for the cavity-EIT resonance. At sufficiently high switching-field intensities, the cavity-EIT feature is strongly distorted, and the system increasingly approaches the response of an empty cavity. Consequently, the difference between the switched and unswitched cavity outputs decreases, resulting in a reduction of the switching contrast. This behavior indicates the existence of an optimum switching-field strength for achieving maximum contrast in the suppression-based switching scheme.

\begin{figure}[htbp]
    
    \includegraphics[width=\columnwidth]{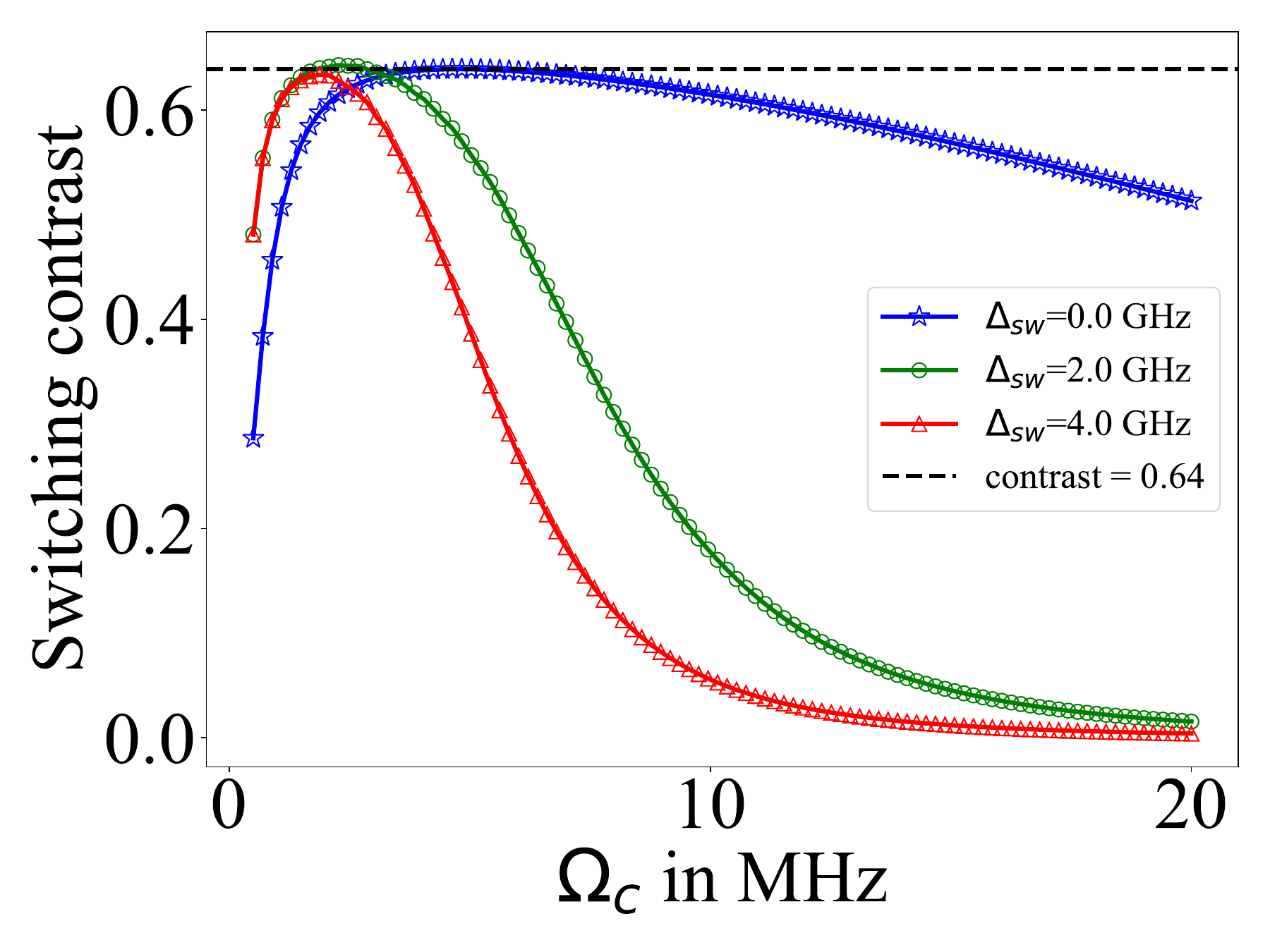}
    \caption{\justifying{
    Dependence of switching contrast on control field Rabi frequency ($\Omega_c$) for 4-level switching. The parameters are taken as $\kappa_1=1.53$ MHz, $\kappa_2=7.85$ kHz,$\kappa_A=0.67$ MHz, $\gamma_{eg}=11.7$ MHz, $\gamma_{e'g}=11.6$ MHz, $g_0=0.54$ MHz, $N=1000$, $\Omega_c=4.35$ MHz, $\Omega_{sw}=75$ MHz.
    }}
    \label{fig:contrast_vs_Omega_c_4lvl}
\end{figure}

% Another important control parameter is the coupling field intensity applied between $\ket{g}$ and $\ket{e}$. The Rabi frequency of the coupling field determines the linewidth of the cavity-EIT spectrum. So increasing $\Omega_c$ makes the linewidth so broader that even shifting of the peak position does not make significant change in the cavity emission intensity at position without applying the switching field. Thus for switching by shifting case switching contrast diminishes when value of $\Omega_c$ increases. For switching by suppression case, we observe the similar behaviour that increasing $\Omega_c$ makes the switching contrast lower but with slow variation with respect to $\Omega_c$ but here the mechanism is different than the shifting case. Here the lowering of contrast actually caused by decoherence induced due to the light shift by large field amplitude like previous case. The switching contrast is also less at very small amplitude of the $\Omega_c$, as with low coupling field amplitude, the quantum fluctuations dominate and lead the cavity-EIT linewidth towards empty cavity behavior.
Another important control parameter is the intensity of the coupling field applied to the $\ket{g}\leftrightarrow\ket{e}$ transition. The corresponding Rabi frequency, $\Omega_c$, determines the linewidth of the cavity-EIT resonance and therefore plays a crucial role in the switching contrast. We have calculated the switching contrast with different values of $\Omega_c$ for both 4-level switching cases (see \autoref{fig:contrast_vs_Omega_c_4lvl}).  
\\

As $\Omega_c$ is increased, the cavity-EIT resonance becomes broader. Consequently, in the resonance-shifting scheme, a given shift of the cavity-EIT resonance produces a progressively smaller change in the cavity emission at a fixed probe frequency. As a result, the switching contrast decreases with increasing $\Omega_c$.
\\

For the suppression-based switching scheme, we observe a similar overall decrease in switching contrast with increasing $\Omega_c$, although the dependence is considerably weaker. The underlying mechanism, however, is different from that of the resonance-shifting scheme. In this case, the reduction in contrast at large $\Omega_c$ is primarily associated with the stronger light-induced shifts and the resulting modification and decoherence of the atomic coherence responsible for cavity-EIT, as discussed previously. Thus, increasing the coupling-field strength beyond the optimum value progressively degrades the contrast of the suppressed state.
\\

Interestingly, the switching contrast is also reduced in the limit of very small $\Omega_c$. When the coupling field is weak, the cavity-EIT resonance becomes increasingly susceptible to quantum fluctuations and decoherence, causing the system to deviate from the ideal cavity-EIT regime and approach the response of an empty cavity. The resulting reduction in the distinction between the EIT and switched states leads to a lower switching contrast. Therefore, both switching schemes exhibit an optimum range of coupling-field strength, where the cavity-EIT feature remains sufficiently well defined while maintaining a large switching contrast.
\\

\begin{figure}[htbp]
    
    \includegraphics[width=\columnwidth]{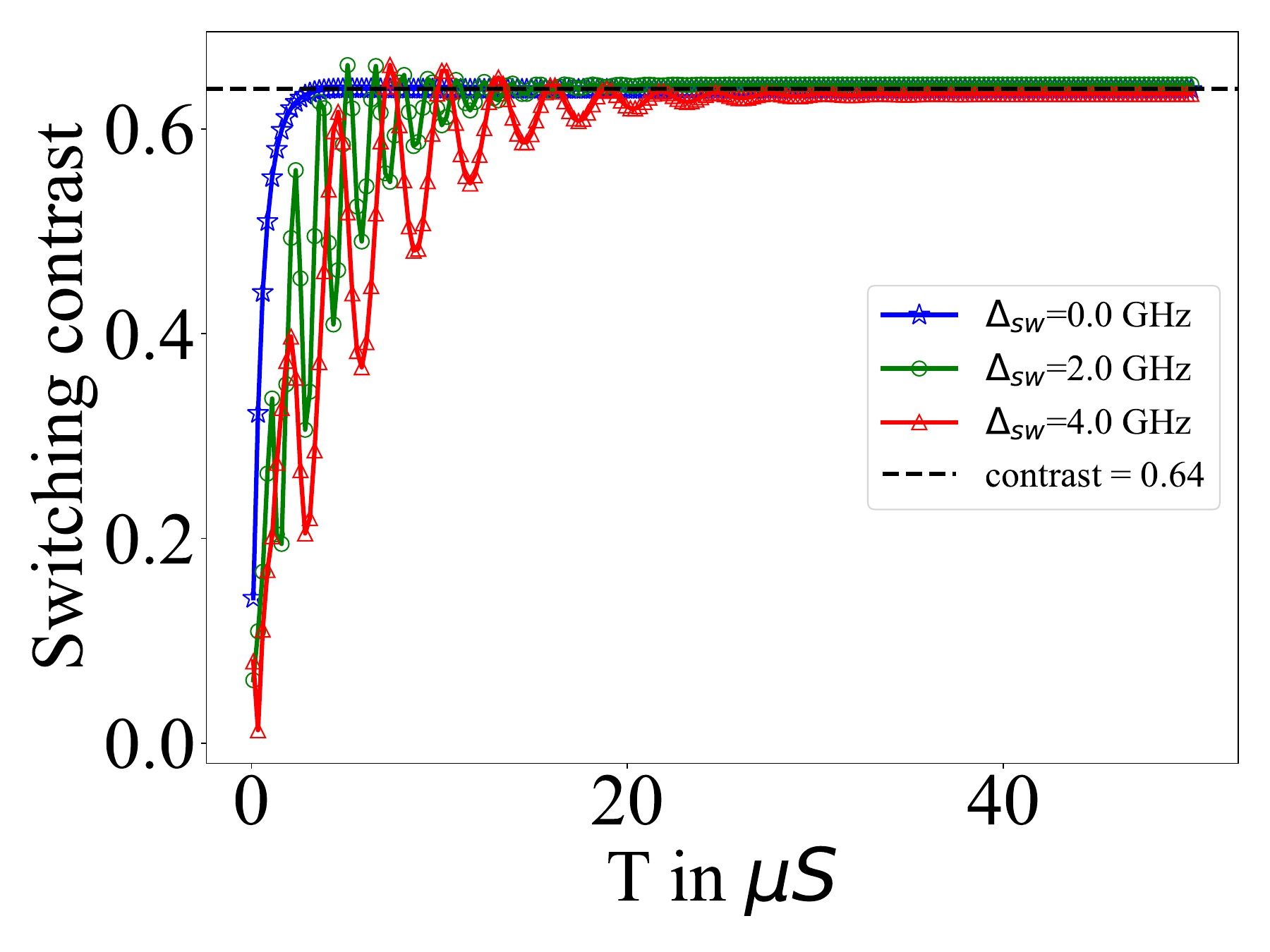}
    \caption{\justifying{
    Switching contrast for different time periods of the applied switching field in the case of 4-level switching. Different parameters with its set values are given by,  $\kappa_1=1.53$ MHz, $\kappa_2=7.85$ kHz,$\kappa_A=0.67$ MHz, $\gamma_{eg}=11.7$ MHz, $\gamma_{e'g}=11.6$ MHz, $g_0=0.54$ MHz, $N=1000$, $\Omega_c=4.35$ MHz, $\omega_{sw}=75$ MHz.
    }}
    \label{fig:contrast_vs_T_4lvl}
\end{figure}

To gain further insight into the switching time and the maximum switching rate that can be achieved while maintaining a high switching efficiency, we calculate the switching contrast for different pulse durations of the switching field. The results are shown in \autoref{fig:contrast_vs_T_4lvl}. For the suppression-based switching scheme, when the switching-field pulse duration is very short, the system does not have sufficient time to respond to the applied field, and the cavity-EIT signal is only partially suppressed. Consequently, the switching contrast remains low. As the pulse duration is increased, the cavity-EIT signal is progressively suppressed, resulting in a corresponding increase in the switching contrast. For sufficiently long pulse durations, the system approaches its steady-state suppressed response, and the switching contrast eventually saturates at its maximum value.
\\

The resonance-shifting scheme exhibits a qualitatively similar dependence on the pulse duration, but with a more pronounced transient behavior. For very short switching pulses, the cavity-EIT resonance does not have sufficient time to shift to its steady-state position, resulting in incomplete switching and a low contrast. As the pulse duration is increased, the resonance progressively evolves toward its steady-state shifted position, leading to an overall increase in the switching contrast. Unlike the suppression scheme, however, the contrast does not necessarily increase monotonically with pulse duration because the cavity-EIT resonance is dynamically evolving during the switching process and may not reach its steady-state position within a given pulse duration. For sufficiently long pulses, the system eventually approaches the steady-state shifted configuration, and the switching contrast converges toward the same saturation value obtained from the steady-state analysis.
\\

These results demonstrate that the pulse duration plays a crucial role in determining the achievable switching contrast. A sufficiently long switching pulse is required for the system to reach the desired switched state, while shorter pulses result in incomplete switching. The dependence of the contrast on pulse duration therefore provides a direct measure of the temporal response of the two switching mechanisms and allows the minimum pulse duration required for high-contrast switching to be determined.
\\

\begin{figure}[htbp]
\begin{subfigure}[t]{\linewidth}
\includegraphics[width=\linewidth]{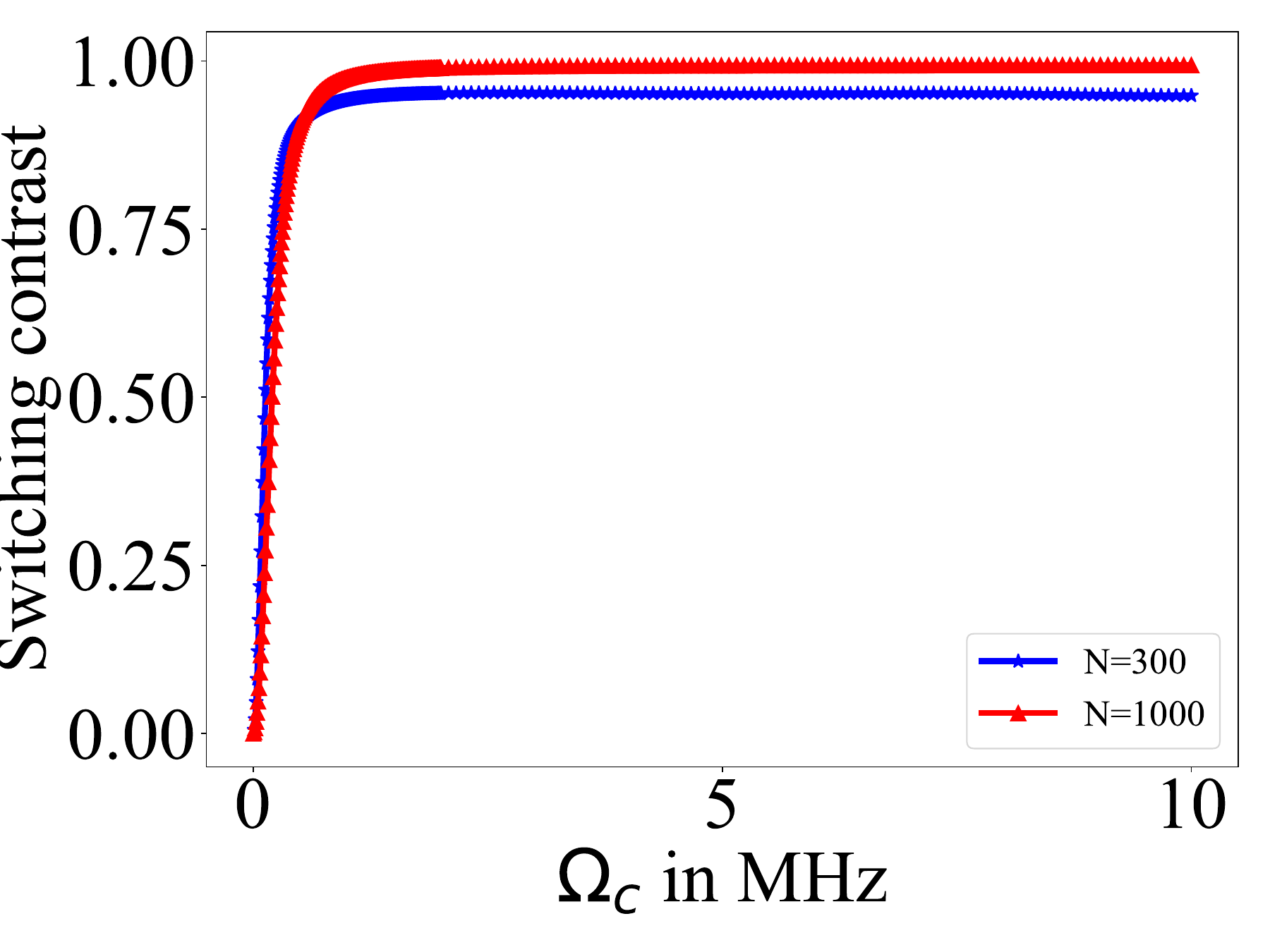}
\caption{}
\label{fig:sw_performance_3lvl_a}
\end{subfigure}
\hfill
\begin{subfigure}[t]{\linewidth}
\includegraphics[width=\linewidth]{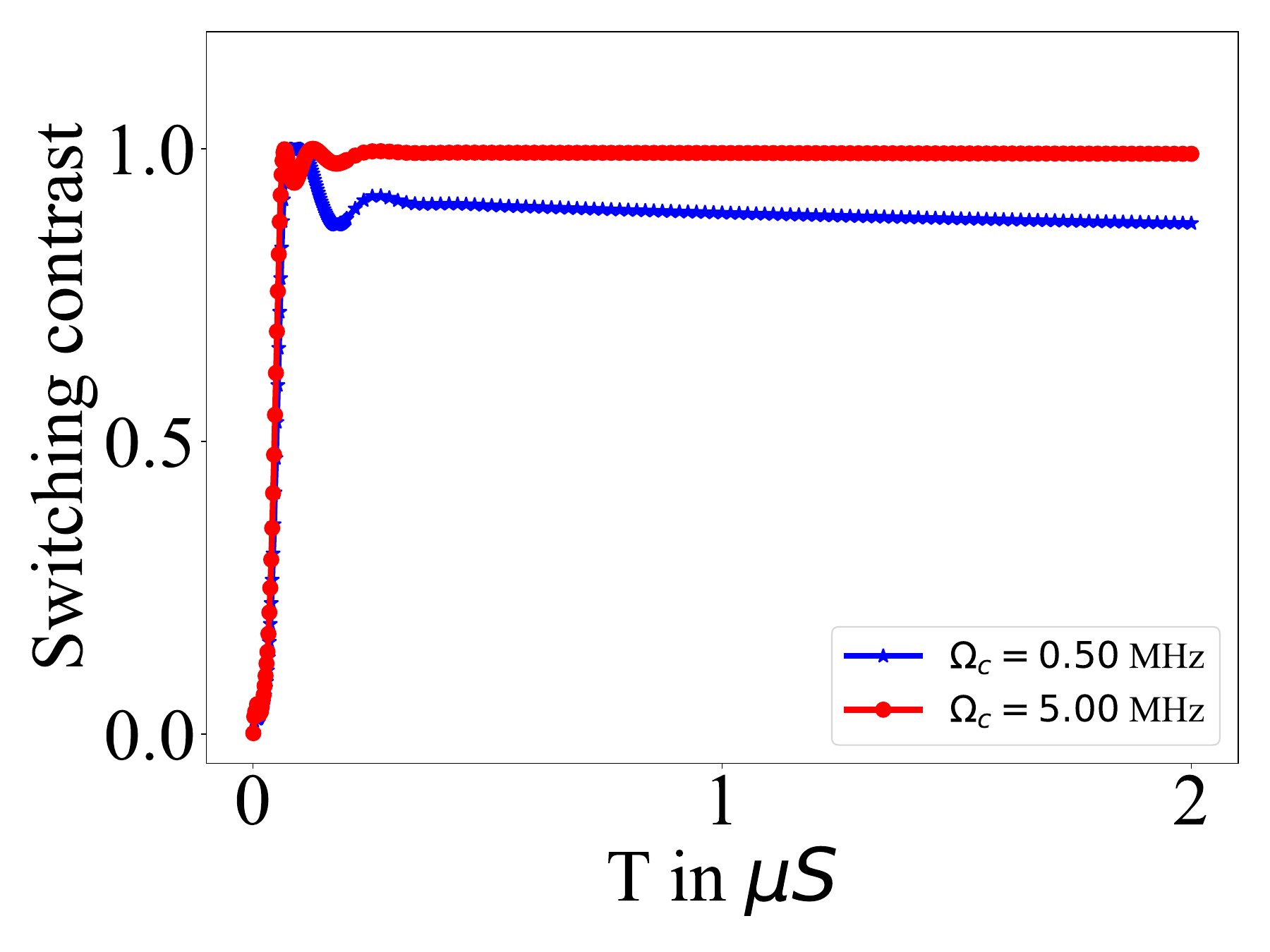}
\caption{}
\label{fig:sw_performance_3lvl_b}
\end{subfigure}
\caption{\justifying {(a) Switching contrast with respect to control field Rabi frequency ($\Omega_c$) for three-level switching. (b) Switching contrast of the 3-level switching for different time periods of the control field pulse. The parameters are taken as $\kappa_1=1.53$ MHz, $\kappa_2=7.85$ kHz,$\kappa_A=0.67$ MHz, $\gamma_{eg}=11.7$ MHz, $\gamma_{eu}=0.84$ MHz,$\gamma_{gu}=1$ kHz, $g_0=0.54$ MHz, $N=1000$.}
}
\label{fig:sw_performance_3lvl}
\end{figure}

% We also analyzed the three level switching performance with the control field Rabi frequency for different coupling strength and pulse time period of switching laser, which is control laser in this case. The results are illustrated in \autoref{fig:sw_performance_3lvl}. 

% In this case increasing $\Omega_c$ gives the same behavior as supression based switching in 4-level system described previously except the fact that contrast does not fade out with increasing laser power and the contrast reaches to 100\% after certain value of $\Omega_c$, which is much lower than the $\Omega_c$ value at maximum switching contrast achievable in 4-level schemes.   This is because here $\Omega_c$ it self switched on and off at a very fast rate instead of applying it for long time. At a region where contrast reaches its saturation maximum, for same value of $\Omega_c$, the contrast decreases with coupling strength (see \autoref{fig:sw_performance_3lvl_a}).

% The timing characteristics illustrated in \autoref{fig:sw_performance_3lvl_b}, shows that with a very short pulse duration and low power the maximum switching contrast can be achievable in this case compared to the 4-level switching. 

We also analyze the switching performance of the three-level system as a function of the control-field Rabi frequency, $\Omega_c$, for different cavity coupling strengths and switching-pulse durations. The corresponding results are presented in \autoref{fig:sw_performance_3lvl}.
\\

In this case, increasing $\Omega_c$ produces a behavior qualitatively similar to that observed for the suppression-based switching in the four-level system discussed previously. However, there is an important difference: the switching contrast does not decrease at large $\Omega_c$. Instead, it approaches $100\%$ beyond a certain coupling-field strength. Moreover, the value of $\Omega_c$ required to reach maximum contrast is substantially smaller than that required to obtain the maximum contrast in the four-level suppression scheme. This difference arises because, in the three-level configuration, the control field itself is directly switched on and off. Thus, the switching operation does not require the system to remain under a strong control field for a sufficiently long duration to establish a steady-state suppressed response. Once the control field is switched off, the EIT condition is destroyed directly, allowing efficient suppression of the cavity-EIT signal even for relatively short switching pulses.
\\

In the regime where the switching contrast reaches its saturation value, we further observe that, for a fixed $\Omega_c$, the contrast increases with increasing cavity coupling strength, as shown in \autoref{fig:sw_performance_3lvl_a}. This dependence reflects the increasing influence of the cavity--atom interaction on the transient dynamics when the system is driven away from the steady-state EIT condition.
\\

The temporal characteristics of the three-level switching are shown in \autoref{fig:sw_performance_3lvl_b}. Remarkably, high switching contrast can be achieved even for very short switching pulses and relatively low control-field power. This behavior highlights the faster transient response of the three-level switching scheme compared with the four-level switching schemes considered above. In particular, because the EIT control field itself is switched off, the suppression of the cavity-EIT signal does not require the system to evolve toward a new steady-state configuration under a continuously applied switching field. Consequently, the three-level scheme can achieve high-contrast switching on a timescale significantly shorter than the characteristic cavity-EIT response time, making it attractive for applications requiring rapid optical gating.

\section{Conclusion}\label{sec:conclusion}
% We have performed a comparative study of different switching schemes theoretically in cavity-EIT based optical switching with respect to the temporal response of switching. The switching contrast is taken as a tool which is calculated from the time dynamic spectrum of cavity output. The contrast is calculated with respect to relevant controlling parameters and compared with different switching schemes to get best performance regime for different switching processes. 

% It is observed that in 4-level-based switching schemes, the maximum achievable contrast with respect to different parameters is similar around 0.65. When we talk about the switching speed, switching by suppression has a much faster response than switching by shifting of resonance. It indicates that if one requires only one of the signals, do not bypass the signal in other modes, suppression-based switching may be useful for achieving high contrast at a low time period of around $5~ \mu s$. 

% For three-level based switching the contrast reaching to unity and switching rate is also very high. Only drawback in this type of switching is as the control field itself is switched on and off here, we are loosing an extra control parameter.

% In conclusion, through the comparative study, we have shown the performance limits of each type of switching with respect to various parameters. This may help to optimize the experiment on cavity-EIT switching according to the particular application. 

We have presented a theoretical comparative study of different cavity-EIT-based all-optical switching schemes in a system of trapped ions coupled to an optical cavity, with particular emphasis on their transient response. The switching performance is characterized in terms of the switching contrast, which is evaluated from the time-dependent cavity output. We systematically investigate the dependence of the switching contrast on the relevant system parameters and compare the temporal response of the different switching schemes to identify their respective performance limits and optimal operating regimes.
\\

For the four-level N-type system, we consider two distinct switching mechanisms: suppression of the cavity-EIT signal and shifting of the cavity-EIT resonance. For both schemes, the maximum achievable switching contrast is found to be approximately $0.65$ over the range of parameters investigated. However, their transient responses are markedly different. Switching through suppression exhibits a significantly faster response than switching through resonance shifting for otherwise identical system parameters. In the suppression scheme, the switching field directly inhibits the cavity-EIT signal, whereas resonance-shift switching requires the cavity-EIT resonance to evolve toward a new spectral position. Consequently, when the application requires only the suppression of a particular cavity-EIT signal, without requiring the signal to be transferred to another spectral mode, suppression-based switching can provide high contrast with relatively short switching pulses, with characteristic pulse durations of approximately $5~\mu\mathrm{s}$ for the parameters considered here.
\\

We have also investigated switching in the corresponding three-level cavity-EIT system, where the control field responsible for establishing the EIT condition is itself switched on and off. In this case, the switching contrast can approach unity, while high contrast can be obtained with substantially shorter switching pulses and lower control-field Rabi frequencies than those required in the four-level schemes. This faster response originates from the direct destruction of the EIT condition when the control field is switched off, rather than from the evolution of the system under an additional switching field. Thus, the three-level configuration offers a significant advantage in terms of switching contrast and speed. Its principal limitation, however, is the lack of an independent switching control parameter, since the field responsible for establishing the cavity-EIT condition is also used to perform the switching operation.
\\

Overall, our comparative analysis demonstrates that the different cavity-EIT switching schemes involve distinct trade-offs between switching contrast, switching speed, and control flexibility. The three-level scheme is particularly attractive when ultrafast, high-contrast ON/OFF switching is the primary requirement. The four-level suppression scheme provides an independently controlled switching channel and can be advantageous when the cavity-EIT operating condition needs to be maintained while the output is selectively suppressed. The resonance-shifting scheme, although requiring higher switching power and longer temporal evolution, provides an additional means of controlling the spectral position of the cavity-EIT response.
\\

These results provide a systematic assessment of the performance limits and operating regimes of the different cavity-EIT switching mechanisms. In particular, by identifying the dependence of the switching contrast and temporal response on the relevant experimental parameters, our analysis provides useful guidance for designing and optimizing experiments aimed at realizing cavity-EIT-based optical switching. The results can therefore serve as a reference for selecting suitable system parameters and switching schemes according to the requirements of a particular experimental implementation and application, including optical gating, optical transistor operation, and frequency-selective photonic signal processing.

% Overall, our comparative analysis demonstrates that the different cavity-EIT switching schemes involve distinct trade-offs between switching contrast, switching speed, and control flexibility. The three-level scheme is particularly attractive when ultrafast, high-contrast ON/OFF switching is the primary requirement. The four-level suppression scheme provides an independently controlled switching channel and can be advantageous when the cavity-EIT operating condition needs to be maintained while the output is selectively suppressed. The resonance-shifting scheme, although requiring higher switching power and longer temporal evolution, provides an additional means of controlling the spectral position of the cavity-EIT response. These results establish the performance limits of the different switching mechanisms and provide useful guidelines for optimizing cavity-EIT-based optical switching according to the requirements of specific applications, including optical gating, optical transistor operation, and controlled photonic signal processing.

\subsection*{Acknowledgments}
Abhijit Kundu gratefully acknowledges financial support from IIT Tirupati, Government of India. Vijay Bhatt acknowledges support from the NQM (National Quantum Mission), Department of Science and Technology, Ministry of Science and Technology, Government of India through Project No. DST/QTC/NQM/QComm/2024/2(G) administered through the IITM CDOT SAMGYNA TECHNOLOGIES FOUNDATION, IIT Madras, Chennai, Tamil Nadu - 600113. Arijit Sharma acknowledges support from the NQM (National Quantum Mission), Department of Science and Technology, Ministry of Science and Technology, Government of India through Project Nos. DST/QTC/NQM/QComm/2024/2(G) and DST/QTC/NQM/QComm/2024/2(C) administered through the IITM CDOT SAMGYNA TECHNOLOGIES FOUNDATION, IIT Madras, Chennai, Tamil Nadu - 600113.

\subsection*{Conflicts of Interest}

The authors declare no conflicts of interest.

% \funding{Sample text inserted for demonstration.}
%     % This section is a list of funder names and grant numbers
    
% \roles{Sample text inserted for demonstration.}
%     % List author names and the contributions made to the article, using terms from the NISO Contributor Roles Taxonomy (CRediT) https://credit.niso.org
    
% \data{Sample text inserted for demonstration.}
%     % For more information on IOP Publishing's research data policy see: https://publishingsupport.iopscience.iop.org/questions/research-data/

\bibliography{bibliography}% Produces the bibliography via BibTeX.

% \appendix
    
% \section{Section 1}\label{appen:a}
% \begin{figure}[htbp]
    
%     \includegraphics[width=\columnwidth]{refl_albert_match.pdf}
%     \caption{\justifying{Comparison between analytical and simulation results of Cavity-EIT-based optical switching. The simulation result is matching with the analytical expression given in \cite{albert2010light}. The parameters are taken as $\kappa_1=1.53$ MHz, $\kappa_2=7.85$ kHz,$\kappa_A=0.67$ MHz, $\gamma_{eg}=11.7$ MHz, $\gamma_{e'g}=11.6$ MHz, $g_0=0.54$ MHz, $N=996$, $\Omega_c=4.35$ MHz, $\omega_{sw}=75$ MHz.
%     }}
%     \label{fig_ana_simu_comp}
% \end{figure}

% \section{Section 2}\label{appen:b}

%\suppdata{}

% \bibliographystyle{iopart-num}
% \bibliography{references}

% \nocite{*}

\end{document}